\documentclass[twocolumn,showpacs,amsmath,amssymb]{revtex4}

\usepackage{graphicx}
\usepackage{dcolumn}
\usepackage{bm}
\usepackage{epsf}
\usepackage{epsfig}
\usepackage{psfig}

\begin{document}

\title{Equivalence between focussed paraxial beams and the quantum harmonic oscillator}

\author{Ole Steuernagel, email: ole@star.herts.ac.uk\\
{\em Department of Physics, Astronomy and Mathematics,
University of Hertfordshire, Hatfield, AL10 9AB, UK}}%

\date{\today}

\begin{abstract}
The paraxial approximation to the scalar Helmholtz equation is
shown to be equivalent to the Schr\"odinger equation for a quantum
harmonic oscillator. This equivalence maps the Gouy-phase of
classical wave optics onto the time coordinate of the quantum
harmonic oscillator and also helps us understand the qualitative
behavior of the field and intensity distributions of focused
optical beams in terms of the amplitude and probability
distributions of quantum harmonic oscillators and vice versa.
\end{abstract}



\maketitle

\section{Introduction}

It is well known that in the limit of short wavelengths, wave
optics can be reduced to ray-optics: Maxwell's equations can be
reduced to the eikonal equations of geometric
optics.\cite{shortwavelength,Born.Wolf.book} Ray optics therefore
is an emergent theory with limited applicability. The ray optical
description of optical image formation is simple and satisfactory
for many purposes. Indeed, if thin lens formulae apply and
complications such as chromatic aberration can be neglected, the
description is so simple that it is standard material in
undergraduate physics education.\cite{textbook.thin.lens} But
optics requires a wave description when the characteristic length
scales of a beam of light become comparable to its wavelength.
Even when the thin lens formula appears to be applicable, a wave
optical description becomes necessary where a focused light beam's
transverse extension shrinks to a point, for example, in all focal
areas of a conventional imaging apparatus. Wave optics limits the
resolution of point-images through aperture dependent point-spread
functions.\cite{Born.Wolf.book,textbook.thin.lens,Saleh.Teich.book,Siegman.book,Haus.buch}

We investigate the fundamental building blocks of general images:
focused, monochromatic beams. Specifically, we consider focused
beams in the paraxial approximation and assume that we can treat
the polarization degrees of freedom separately so that we can
apply the paraxial approximation to the scalar Helmholtz equation.
Since a general image is described by a suitable mixture of
coherent beams (of different frequencies, polarization, etc.) our
considerations are of some generality. Beams that are not too
strongly focused can typically be decomposed into a complete
orthonormal set of modes that have a simple analytical
description. The most widely used are the Hermite-Gaussian
transverse electromagnetic modes (TEM-modes), which we will
introduce in Sec.~\ref{gauss.beams}. These modes are essentially
suitably rescaled wave functions of a harmonic oscillator
multiplied by phase factors arising from the specific geometry of
the beam, see
Eq.~(\ref{HG.TEM}).\cite{Saleh.Teich.book,Siegman.book,Haus.buch}
This connection between coherent optical beams and harmonic
oscillators is known, but it seems to be little appreciated that
the two systems are isomorphic (but see
Ref.~\onlinecite{Nienhuis}). The amplitude of a focused paraxial
beam corresponds to the amplitude of a quantum harmonic
oscillator, the former's intensity to the probability distribution
of the latter, and the Gouy-phase of optics\cite{FengOL01} assumes
the meaning of time for the quantum harmonic oscillator.

We derive this equivalence by a transformation of the paraxial
wave equation of optics into the Schr\"odinger equation for the
quantum harmonic oscillator in Sec.~\ref{eq_trafo}. We then give a
few illustrative examples for this equivalence in
Sec.~\ref{examples}, which will allow us to discuss some
counterintuitive aspects of optical beam behavior.

\section{Hermite-Gaussian Beams: TEM-modes}\label{gauss.beams}

General monochromatic TEM-beam modes in the paraxial approximation
are solutions of the associated scalar Helmholtz equation for the
electro-magnetic vector potential $\bf A$ propagating in vacuum at
speed $c$ (with the dispersion relation,
$\omega=ck$)\cite{Haus.buch}
\begin{equation}
\nabla^2 {\bf A}({\bf r},t) - \frac{1}{c^2} \frac{{\partial}^2
{\bf A}({\bf r},t) }{\partial t^2} = 0, \label{Helmholtz}
\end{equation}
where ${\bf r} =(x,y,z)$ and $t$ describe position and time
coordinates. A monochromatic beam with wave number $k$ travelling
in the positive $z$-direction and linearly $x$-polarized is
described by the vector potential ${\bf A} = (A_x, A_y, A_z)$
whose only non-zero component is $A_x$ with\cite{Haus.buch}
\begin{equation}
A_x({\bf r},t;k) = \psi({\bf r})\; e^{i(k z - \omega t)}.
\label{Ax}
\end{equation}
If we insert this ansatz into the Helmholtz equation
(\ref{Helmholtz}) and factor out the common plane wave phase
factor $e^{i(k z - \omega t)}$, we obtain a differential equation
for the field envelope $\psi ({\bf r})$ alone:
\begin{subequations}
\label{paraxHelmholtz}
\begin{eqnarray}
0 & = & \frac{{\partial}^2 \psi({\bf r}) }{\partial x^2} +
\frac{{\partial}^2 \psi({\bf r}) }{\partial y^2} +
\frac{{\partial}^2 \psi({\bf r}) }{\partial z^2} + 2 i k
\frac{{\partial} \psi({\bf r}) }{\partial z}
\\
& \approx & \frac{{\partial}^2 \psi({\bf r}) }{\partial x^2} +
\frac{{\partial}^2 \psi({\bf r}) }{\partial y^2} + 2 i k
\frac{{\partial} \psi({\bf r}) }{\partial z}. \label{3b}
\end{eqnarray}
\end{subequations}
Equation~(\ref{3b}) represents the paraxial approximation; it is
based on the observation that, for beams that are not too strongly
focused (several wavelengths across), the $z$-dependence of the
field derivative is primarily due to the plane wave factor $2ik$,
which allows us to discard the dependence on $\partial
\psi/\partial z$. When we perform the second derivative, the term
$\partial^2 \psi/\partial z^2$ is consequently assumed to be
negligible.\cite{Haus.buch}

The paraxial approximation significantly simplifies our task
because it has the familiar transverse electromagnetic
TEM$_{mn}$-modes $\psi_{mn}$ as
solutions.\cite{Saleh.Teich.book,Siegman.book,Haus.buch} They
contain products of Gaussians and Hermite polynomials in the
transverse beam coordinates $x$ and $y$, that is, the familiar
harmonic oscillator wave functions $\varphi_m(\xi) =
H_m(\xi)\exp(-\xi^2/2)/\sqrt{2^m m! \sqrt{\pi} } $,
$(m=0,1,2,\ldots),$ and various phase
factors,\cite{Saleh.Teich.book,Siegman.book,Haus.buch} namely
\begin{eqnarray}
\psi_{mn}({\bf r}) =\frac{w_0}{w(z)}
\varphi_m(\frac{\sqrt{2}\,x}{w(z)})\,
\varphi_n(\frac{\sqrt{2}\,y}{w(z)}) \nonumber \\
\times \; e^{\frac{i k}{2 R(z)}(x^2+y^2)} \; e^{-i (m+n+1)
\phi(z)} . \label{HG.TEM}
\end{eqnarray}
The relation that links the minimal beam radius $w_0$ with the
Rayleigh range $b$ via the light's wavelength $\lambda$ and wave
number $k$ is given by $w_0 = \sqrt{2b/k} = \sqrt{\lambda b/\pi}$.
The beam radius (where the intensity has dropped to $1/e^2$ of the
central intensity) at the distance $z$ from the beam waist obeys
$w(z)= \sqrt{w_0^2 (1 + z^2/b^2)}$, and for large $z$ implies the
expected amplitude decay of a free wave $\propto 1/|z|$ with a
far-field opening angle $\arctan(\lambda/(\pi w_0))$.

The corresponding wave front curvature is described by the radius
$R(z)= (z^2 + b^2)/z$, and the longitudinal Gouy-phase
shift\cite{Siegman.book,Haus.buch} follows
\begin{eqnarray}
\phi(z)=\arctan(z/b) \; , \label{gouy.phase}
\end{eqnarray}
which varies most strongly at the beam's focus. Note that the
associated Gouy-phase factor $e^{-i (m+n+1) \phi(z)}$ depends on
the order $n$ and $m$ of the mode functions. Different modes
therefore show relative dephasing or mode-dispersion, particularly
near the focus.

The vector potential $A_x$ of Eq.~(\ref{Ax}) describes a beam
travelling in the positive $z$-direction (${\bf k}=k\hat{\bf z}$)
and yields an electric field that is polarized in the
$x$-direction with a small contribution in the $z$-direction due
to the tilt of wave fronts off the beam axis.\cite{Haus.buch}
According to Maxwell's equations in the paraxial approximation,
that is, neglecting transverse derivatives, we find for the
electric field vector,\cite{Haus.buch} ($\hat{\bf x}, \hat{\bf y},
\hat{\bf z}$ are the unit-vectors and Re stands for the real part)
\begin{equation}
{\bf E}({\bf r},t;{\bf k}) = \mbox{Re} \big\{ [ \hat{\bf x} \omega
\psi({\bf r}) + \hat{\bf z} ic  \frac{\partial \psi({\bf
r})}{\partial x} ] e^{i(k z - \omega t)} \big\}.
\label{true.E.field}
\end{equation}
The associated instantaneous electrical intensity distribution
is\cite{Haus.buch}
\begin{equation}
I({\bf r},t) = \frac{\epsilon_0}{2} {\bf E}({\bf r},t)^2 \approx
\frac{\epsilon_0}{2} \; \omega^2 \mbox{Re} \{A_x({\bf r},t)\}^2 .
\label{Intensity}
\end{equation}

\section{Transformation to Harmonic Oscillator Schr\"odinger Equation
\label{eq_trafo}}
We have derived the paraxial approximation of the wave equation in
Eq.~(\ref{paraxHelmholtz}) and found that the solutions, except
for some phase factors and a coordinate rescaling in the arguments
of the solutions, look like those of the quantum harmonic
oscillators. We shall therefore now ``undo'' these various factors
and transform the paraxial wave equation into Schr\"odinger's
equation to establish the isomorphism between both systems.
%
\begin{figure}[h t]
\epsfverbosetrue \epsfxsize=3.4in \epsfysize=1.6in
\epsffile[95 135 505 630]{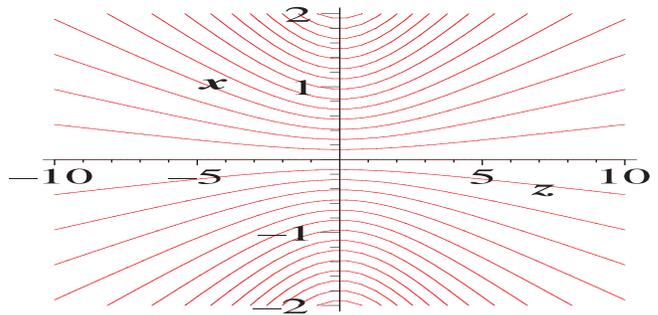}
\caption{Plot of new coordinate lines $\xi = \mbox{constant}$ in
the $xz$-plane. \label{fig1}}
\end{figure}
%

Based on the shape of the solutions (\ref{HG.TEM}), we apply the
following coordinate transformations $[x,y,z]\mapsto [\xi,\eta,
\tau]$, which establishes new coordinate lines $\xi$ and $\eta$
that follow the focused beams' hyperbolic flow lines (see
Fig.~\ref{fig1})
\begin{equation}
x = \xi \frac{ w(z)}{\sqrt{2}}, \quad y = \eta \frac{
w(z)}{\sqrt{2}}.
\end{equation}
Because of Eq.~(\ref{gouy.phase}), we let
\begin{equation}
z = b\tan(\tau), \label{coord.trafo}
\end{equation}
where $\tau$ plays the role of time. If we use these new
coordinates and the ansatz $\psi (x,y,z) = \Psi (\xi,\eta;\tau)
e^{\frac{i k}{2 R(z)}(x^2+y^2)}/w(z)$ in the paraxial wave
equation (\ref{paraxHelmholtz}), we find after a lengthy but
straightforward calculation
\begin{equation}
\Big[ -{\frac {\partial ^{2}}{\partial {\eta}^{2}}} - {\frac
{\partial ^{2}}{\partial {\xi}^{2}}} + {\eta}^{2} + {\xi}^{2} -
2\,i{\frac {\partial }{\partial \tau} } \Big] \Psi(\xi,\eta;\tau)
= 0, \label{schroedinger_equation}
\end{equation}
which is the Schr\"odinger equation for the harmonic oscillator.
The units are $\hbar = m = \Omega =1$, where $m$ is the mass and
$\Omega= 2\pi/T$ is the resonance frequency of the harmonic
oscillator. We have established the equivalence between focused
beams and harmonic quantum oscillators.

Note that the $z$-coordinate $z =(-\infty,\infty)$ in the optical
case was transformed to time $\tau=(-\pi/2,\pi/2)$ in the harmonic
oscillator case, whereas the role of time in the optical case has
no analog in the harmonic oscillator case because we had to factor
out the plane wave phase factor $e^{i(kx-\omega t)}$. In other
words, the isomorphism we established links the beam's envelope
amplitude $\psi (x,y,z)$ with a two-dimensional harmonic
oscillator wave function $\Psi (\xi,\eta;\tau)$.

%
\section{Examples: Single-mode and Two-mode Profiles}\label{examples}
%
%
\begin{figure}[t]
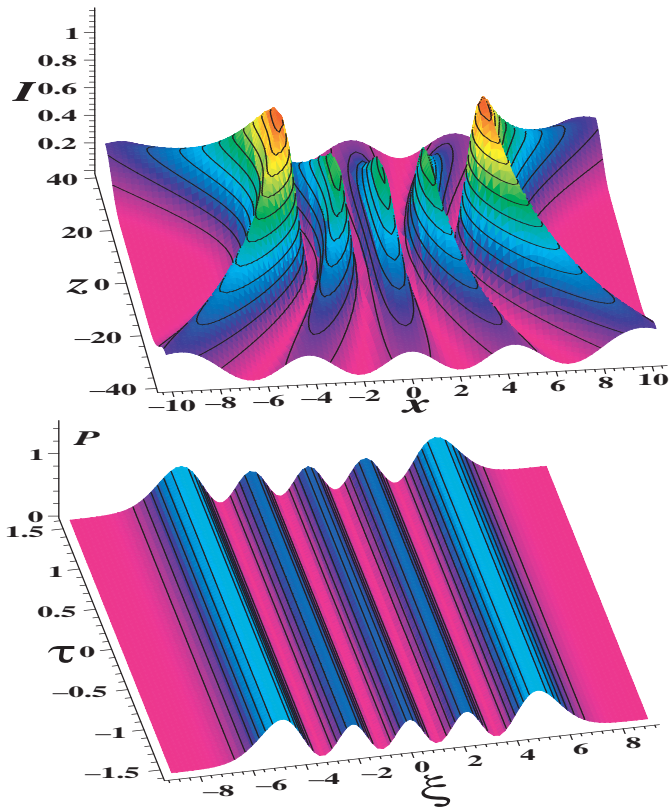

\centering
\includegraphics[width=3.4in,height=2.0in,bb= 065 100 490 580]{fig2a.eps}
\\
\includegraphics[width=3.4in,height=2.0in,bb= 90 105 495 580]{fig2b.eps}
\caption{Top: Contour plot of the time-averaged intensity
$I(x,0,z)$ of a field with mode structure TEM${}_{40}$, namely,
$\varphi_4(\sqrt{2}x/w(z)) \varphi_0(\sqrt{2}y/w(z))$, around the
beam focus $(0,0,0)$; Rayleigh length $b= 20 \lambda$. In this and
all other figures of optical beams the intensity is in arbitrary
units and the coordinates are in units of $\lambda$. Bottom:
Contour plot of the evolution of the probability distribution of a
one-dimensional quantum harmonic oscillator with wave function
$\Psi(\xi) = \varphi_4(\xi)$. \label{fig2}}
\end{figure}
\begin{figure}[h t]
\centering
\includegraphics[width=3.4in,height=2.0in,bb= 60 95 495 600]{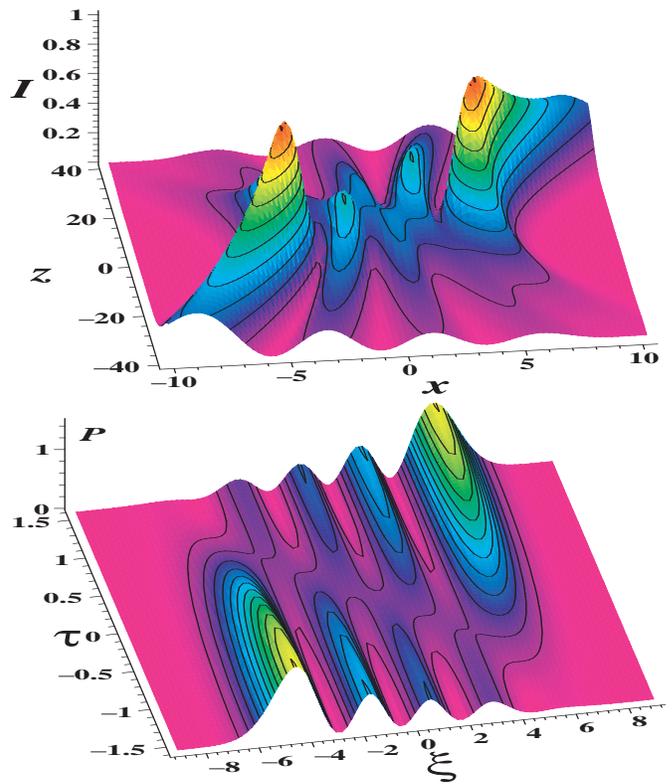}
\\
\includegraphics[width=3.4in,height=2.0in,bb= 85 100 490 605]{fig3b.eps}
\caption{Average intensity $I(x,0,z)$ of a field with mode
structure $[\varphi_4(\sqrt{2}x/w_0) - i \varphi_3(\sqrt{2}x/w_0)]
\varphi_0(\sqrt{2}y/w_0)$ at the beam focus $z=0$; $b=20 \lambda$.
Bottom: Contour plot of the evolution of the non-normalized
probability distribution of a one-dimensional quantum harmonic
oscillator with wave function $\Psi(\xi;0) = \varphi_4(\xi)- i
\varphi_3(\xi)$. \label{fig3}}
\end{figure}

%
For simplicity, we will now consider one-dimen\-sio\-nal beam
patterns. We choose the $y$-mode to be the zero-mode, and thus
concentrate on the behavior in the transverse $x$-direction only,
that is, we investigate the beams in the $xz$-plane. For easy
comparison, we depict focused beams with a beam parameter of $b=
20 \lambda$. Our first example is that of a single
TEM$_{40}$-beam, plotted in Fig.~\ref{fig2}; we next consider a
more interesting case, a two-mode super\-position of TEM-modes 30
and 40 (see Fig.~\ref{fig3}).
%
%

For the two-mode superposition in Fig.~\ref{fig3} we can clearly
see the interference between modes $\varphi_3$ and $\varphi_4$ at
work and how it leads to the redistribution of the intensity from
one beam edge to the other, thus inverting the transverse beam
profile about the beam axis. This intensity redistribution is
ultimately due to the mode-dispersive effect of the Gouy-phase
factor $e^{i(m+n+1)\phi}$ in the optical case and the time phase
factor $e^{i (m + 1/2 + n + 1/2) \Omega t}$ for the
two-dimensional harmonic oscillator. Note that a beam that is
focused only in one transversal direction (using a cylindrical
lens rather than a spherical one) corresponds to a one-dimensional
harmonic oscillator,\cite{FengOL01} because in this case the
Gouy-phase factor is $e^{i(m+1/2)\phi}$ and corresponds to a
one-dimensional harmonic oscillator phase factor $e^{i (m + 1/2)
\Omega t}$. Clearly, both systems evolve through half an
oscillation period ($t=-T/4,\ldots,T/4$. (Remember, the Gouy-phase
(\ref{gouy.phase}) varies from $-\pi/2$ to $\pi/2$.)

We know from ray optics that images become inverted at the focus.
Here we see how this image inversion arises through the action of
the Gouy-phase. In the language of the harmonic oscillator, it
corresponds to half an oscillation $[-\pi/2,\pi/2]$ or swinging to
the ``other'' side.
%
\begin{figure}[h t]
\centering
\includegraphics[width=3.4in,height=3.2in,bb= 55 80 490 645]{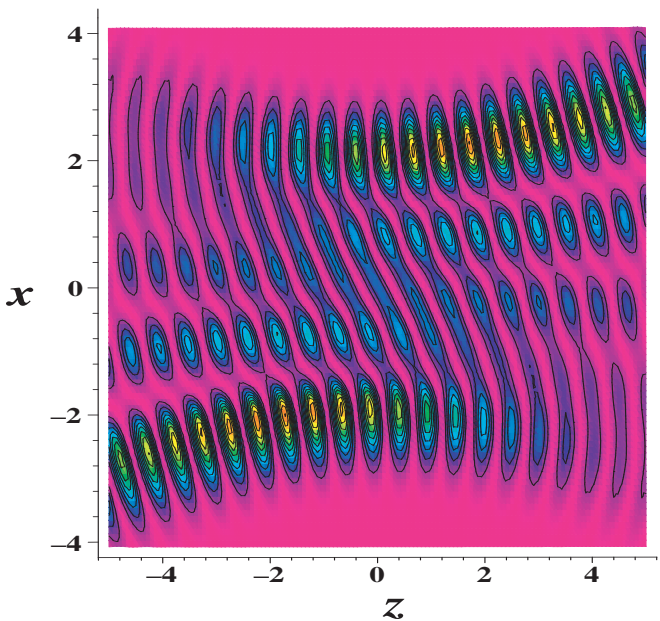}
\caption{Instantaneous intensity $I(x,0,z;0)$ of a field with mode
structure $[\varphi_4(\sqrt{2}x/w_0) - i \varphi_3(\sqrt{2}x/w_0)]
\varphi_0(\sqrt{2}y/w_0)$ at the beam focus $z=0$. For clarity, a
smaller beam parameter of $b=5 \lambda$ was chosen.
\label{fig3i4_wavefronts}}
\end{figure}
%

To gain further insight, we plot the beam's instantaneous
intensity distribution $I(x,0,z;0)$ in
Fig.~\ref{fig3i4_wavefronts}. It can clearly be seen that the
relative phase of $90^\circ$ between modes 4 and 3 tilts the wave
fronts, and the action of the Gouy-phase ``transports'' the beam's
envelope across the focus.
%
\begin{figure}[h t]
\centering
\includegraphics[width=3.4in,height=2.0in,bb= 60 115 505 550]{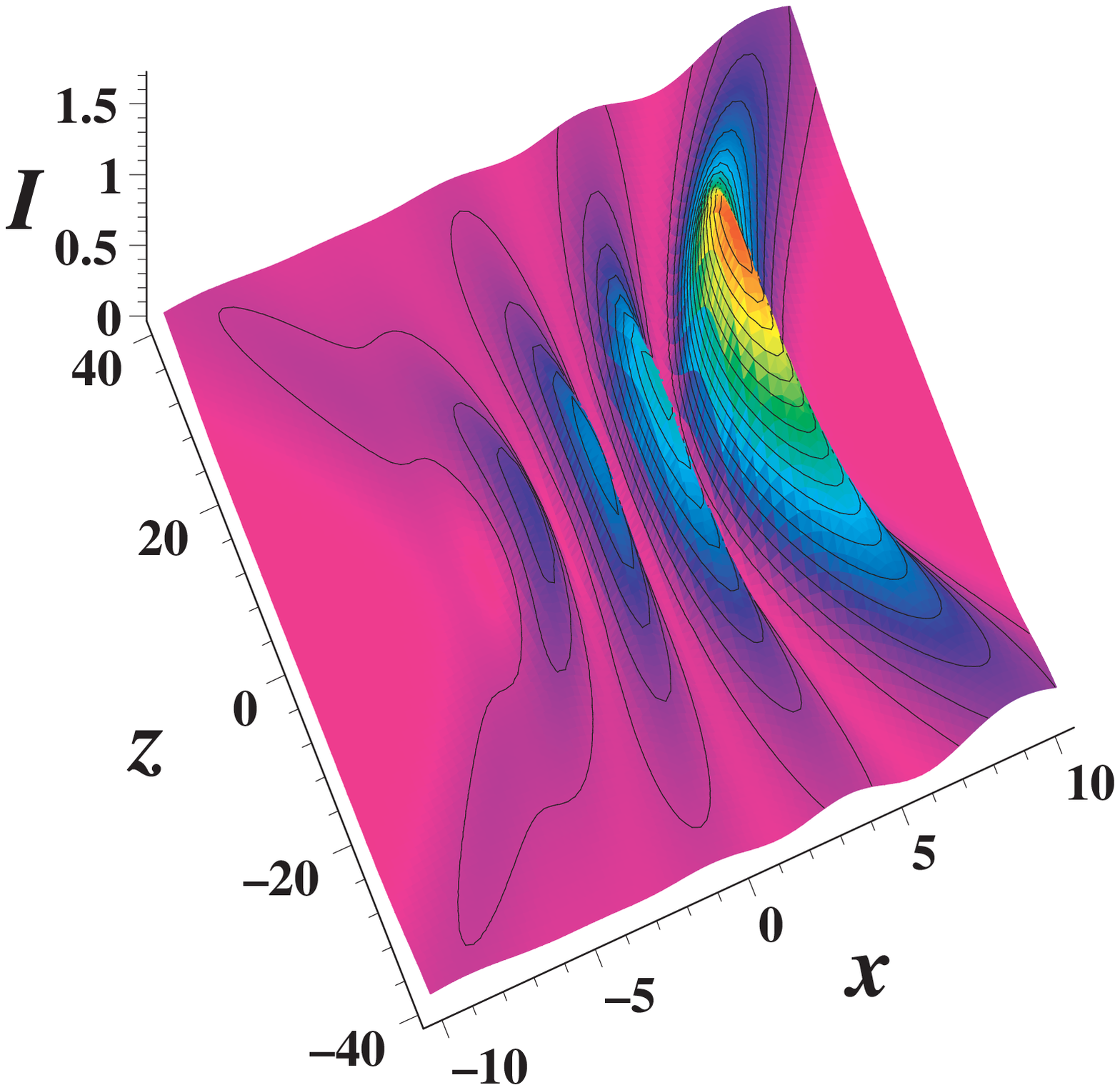}
\\
\includegraphics[width=3.4in,height=2.0in,bb= 80 95 490 580]{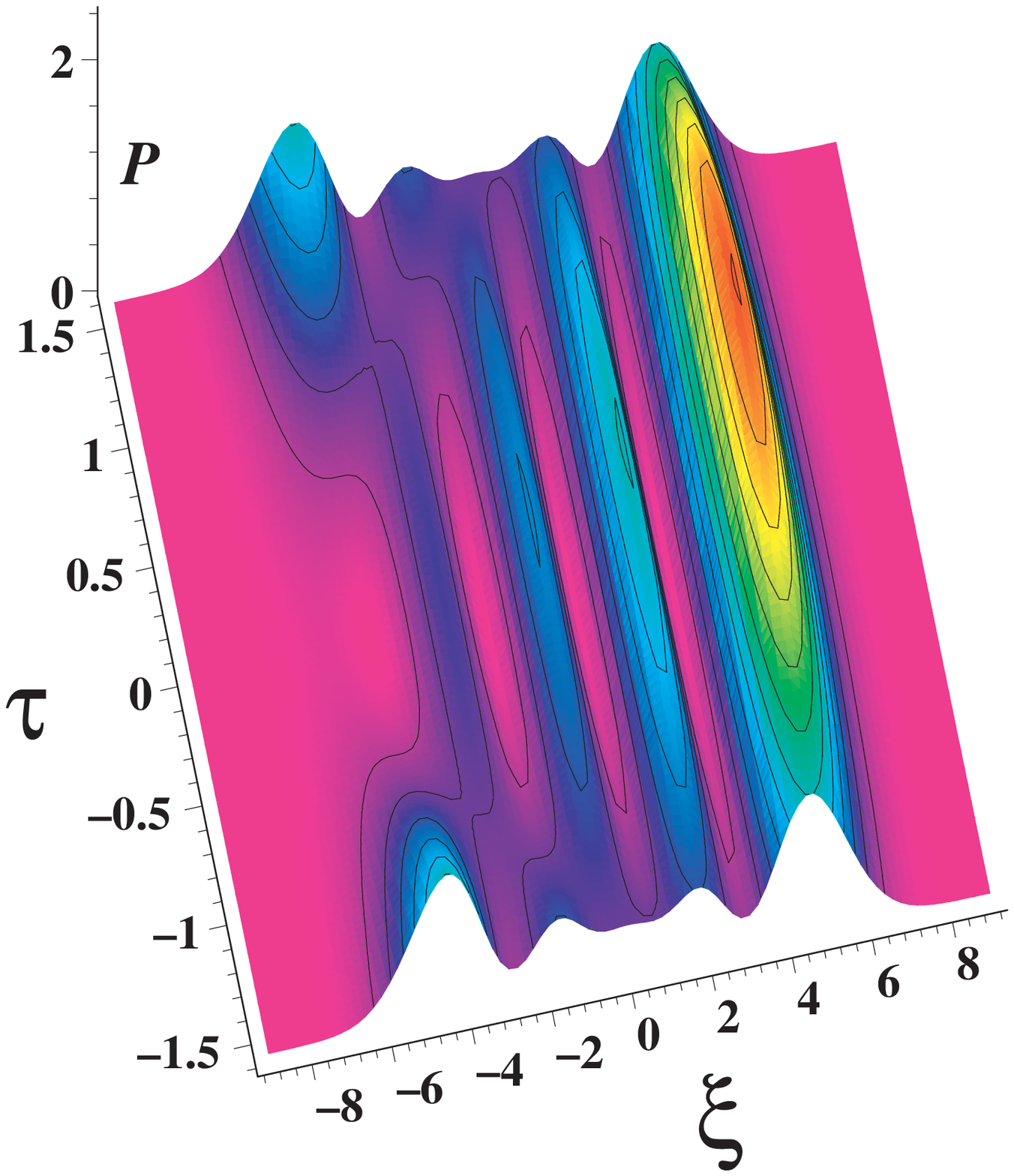}
\caption{Average intensity $I(x,0,z)$ of a field with mode
structure $[\varphi_4(\sqrt{2}x/w_0) + \varphi_3(\sqrt{2}x/w_0)]
\varphi_0(\sqrt{2}y/w_0)$ at the beam focus $z=0$; $b=20 \lambda$.
Bottom: Contour plot of the evolution of the non-normalized
probability distribution of a quantum harmonic oscillator with
wave function $\Psi(\xi;0) = \varphi_4(\xi) + \varphi_3(\xi)$.
\label{fig_34}}
\end{figure}
%

Similarly to the case discussed in Fig.~\ref{fig3}, but at first
seemingly less intuitively, we can also have a scenario where the
harmonic oscillator starts out with an average position at the
beam center, then swings up to one side and falls back to the
center. In wave optical terms this scenario corresponds to a field
with a symmetric far-field distribution that is asymmetric at the
focus and becomes symmetric in the far-field on the other side,
this is illustrated in Fig.~\ref{fig_34}.
%
\begin{figure}[h t]
\centering
\includegraphics[width=3.4in,height=2.0in,bb= 60 105 505 560]{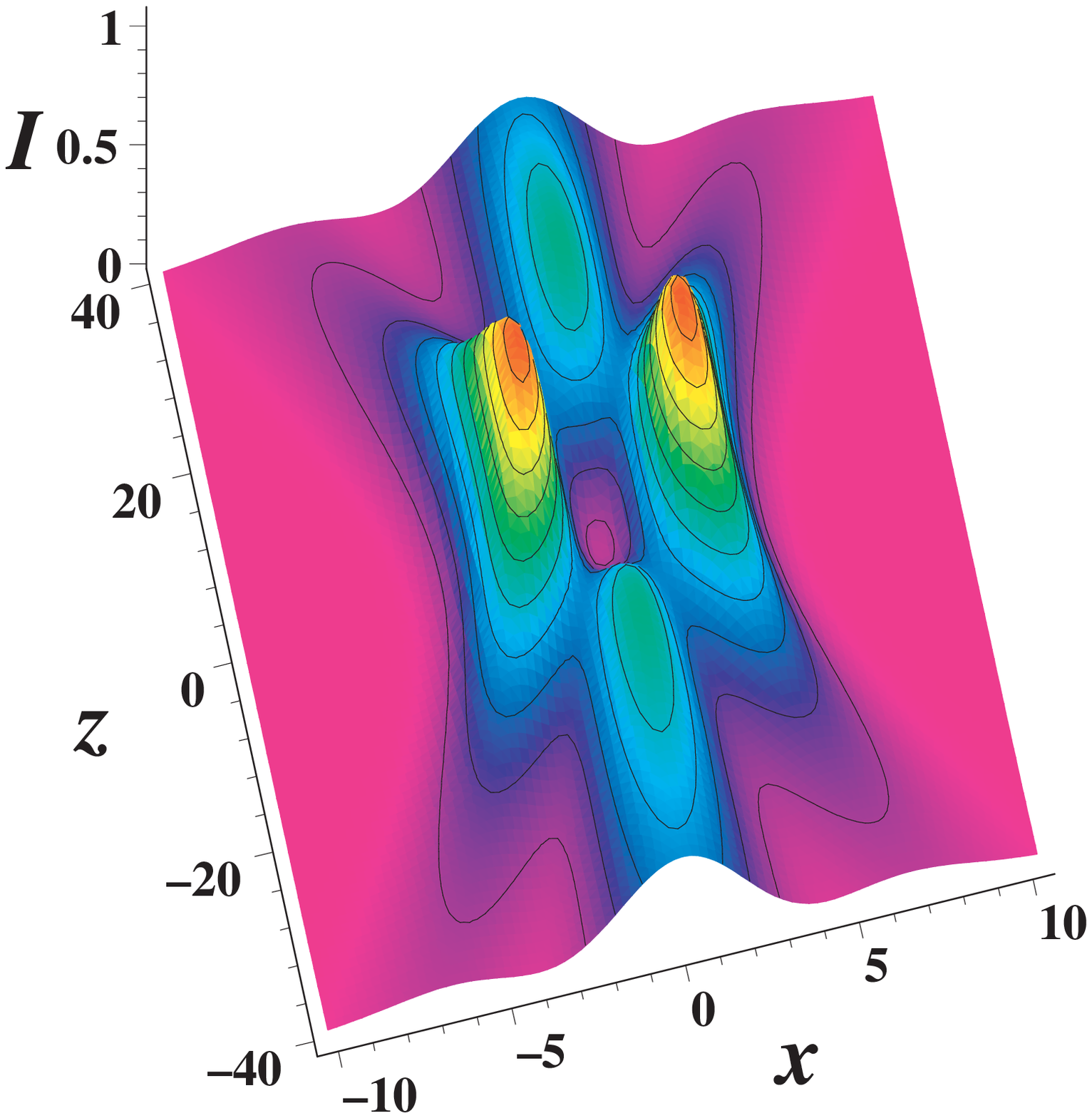}
\\
\includegraphics[width=3.4in,height=2.0in,bb= 80 100 500 600]{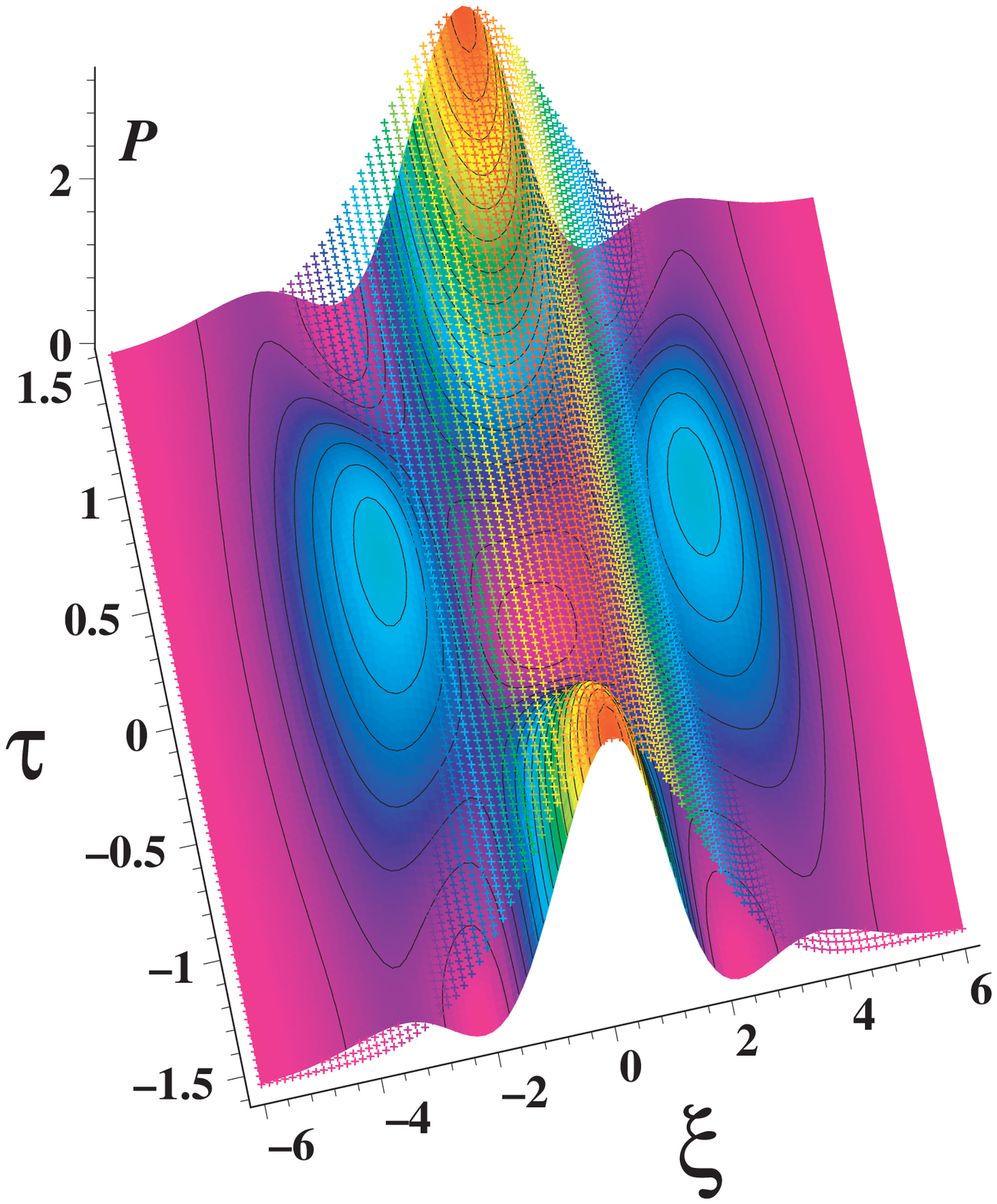}
\caption{Average intensity $I(x,0,z)$ of a field with mode
structure $[\varphi_0(\sqrt{2}x/w_0) + \varphi_2(\sqrt{2}x/w_0)]
\varphi_0(\sqrt{2}y/w_0)$ at the beam focus $z=0$; $b=20 \lambda$.
Bottom: Contour plot of the evolution of the non-normalized
probability distribution of a quantum harmonic oscillator with
wave function $\Psi(\xi;0) = \varphi_0(\xi)+ \varphi_2(\xi)$. For
comparison this distribution is overlayed by the non-normalized
distribution for the ground state wave function $\varphi_0(\xi)$
alone. It shows that the superposition state $\Psi$ is very
tightly confined in the $\xi$-direction, see text. \label{fig_02}}
\end{figure}
%

It also is possible to create a beam that, in oscillator language,
starts out in the center and swings to either side to fall back to
the center forming an annulus with a dark center at the beam's
focus (see Fig.~\ref{fig_02}). It also can be viewed as a beam
that is overly focused in the far field. Then, according to
Heisenberg's uncertainty principle $\Delta(x) \Delta(p_x)\geq
\hbar /2$, such a spatially strongly localized harmonic oscillator
carries a large momentum uncertainty. The bottom of
Fig.~\ref{fig_02} displays the harmonic oscillator ground state
together with the superposition of modes 0 and 2. Clearly the
superposition is spatially more confined than the ground state,
and its Fourier transform reveals a large spread, namely, a
central peak and two momentum peaks that are displaced from zero.
In other words, this superposition describes a harmonic oscillator
that ``explodes'' sideways and then swings back to the center. The
seemingly counterintuitive formation of a dark center at the focus
becomes easy to understand. Its two-dimensional analog is the
optical bottle beam observed very recently.\cite{optical-bottles}
The latter is formed from a superposition of Laguerre-Gaussian
modes 0 and 2, and thus carries basically the same structure in a
cylindrical geometry as our example of Fig.~\ref{fig_02} in the
$xz$-plane.

\begin{figure}
\epsfverbosetrue \epsfxsize=3.4in \epsfysize=1.3in
\epsffile[60 180 700 365]{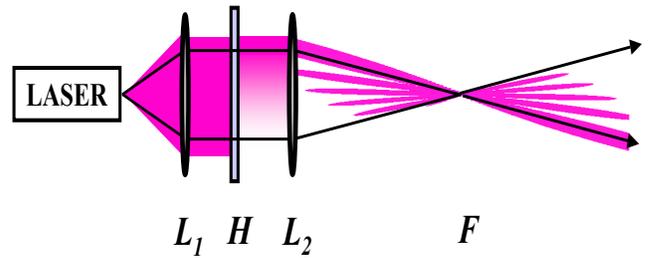}
\caption{Sketch of possible implementation using a laser, two
lenses $L_1$ and $L_2$, and a suitable hologram $H$. The hologram
modulation of the laser beam is mapped into the focal region $F$
of lens $L_2$. Formally, the transverse field profile in its focal
plane is the Fourier-transform of the hologram-pattern.
\label{setup.sketch}}
\end{figure}

A possible technical implementation of the patterns discussed here
is sketched in Fig.~\ref{setup.sketch} and has recently been
demonstrated experimentally.\cite{ole.glasgow} For further details
on experimental implementations using computer-generated
holograms, consult
Refs.~\onlinecite{optical-bottles,ole.glasgow,computer.hologram}.
The field profiles sketched here should be useful for optical
manipulation techniques as well.\cite{ole_jopa05}
%
\nopagebreak
%

\section{Conclusion \label{conclusion}}
Focused beams cannot be described by ray optics in their focal
region. Therefore an intuitive understanding of the wave optical
behavior of a coherent, focused beam with transverse intensity
modulation is desirable. It is shown that the paraxial wave
equation is equivalent to the Schr\"odinger equation of a
two-dimensional harmonic oscillator. This equivalence is used to
show how the passage of a beam through its focus can be understood
in terms of the behavior of a harmonic oscillator evolving through
half a period.

In particular, the inversion of a focused beam's far-field
intensity pattern about the beam axis, upon passage through the
focus, is described in terms of a harmonic oscillator swinging to
the ``other'' side, and the formation of a bottle beam is
explained by the behavior of a constricted beam pattern that
corresponds to a harmonic oscillator fanning outward and
recollapsing.

\begin{acknowledgments}
I wish to thank Mark Dennis for pointing out that the formal
equivalence between paraxial focused beams and quantum harmonic
oscillators is not well known. I also thank Paul Kaye, Joseph
Ulanowski, Jon Marangos, Ed Hinds, Miles Padgett, Johannes
Courtial, and Eric Yao for stimulating discussions, and a referee
for some clarifications.
\end{acknowledgments}

\end{document}